\documentclass[a4paper]{jpconf}
\usepackage{graphicx}
\begin{document}
\title{Modelling the light curves of {\it Fermi} LAT millisecond pulsars}

\author{C Venter$^1$, T J Johnson$^{2,3}$, A K Harding$^4$ and J E Grove$^2$}
\address{
$^1$ Centre for Space Research, North-West University, Potchefstroom Campus, Private Bag X6001, Potchefstroom 2520, South Africa\\
$^2$ High-Energy Space Environment Branch, Naval Research Laboratory, Washington, DC 20375, USA\\
$^3$ National Research Council Research Associate, National Academy of Sciences, Washington, DC 20001\\
$^4$ Astrophysics Science Division, NASA Goddard Space Flight Center, Greenbelt, MD 20771, USA\\
}

\begin{abstract}
We modelled the radio and $\gamma$-ray light curves of millisecond pulsars using outer gap, two-pole caustic, low-altitude slot gap, and pair-starved polar cap geometric models, combined with a semi-empirical conal radio model. We find that no model fits all cases, with the outer gap and two-pole caustic models providing best fits for comparable numbers of millisecond pulsar light curves. We find a broad distribution of best-fit inclination angles as well as a clustering at large observer angles. The outer gap model furthermore seems to require relatively larger inclination angles, while the two-pole caustic model hints at an inverse trend between inclination angle and pulsar spin-down luminosity. 
\end{abstract}

\section{Introduction: why millisecond pulsars are special}
There are 40 millisecond pulsars (MSPs) in the second pulsar catalogue (2PC)~\cite{2PC} of the {\it Fermi} Large Area Telescope (LAT)~\cite{Atwood09}. These represented roughly a third of the then-known $\gamma$-ray pulsar population\footnote{The newest MSP fraction is now in excess of 40\%. See https://confluence.slac.stanford.edu/display/GLAMCOG/\\ Public+List+of+LAT-Detected+Gamma-Ray+Pulsars}. Their uniqueness is related to their relatively small magnetospheres. This is due to their small periods $P$, which determine their so-called light cylinder radius ($R_{\rm LC} = Pc/2\pi$) where the corotation speed equals the speed of light, acting as a boundary characterising the magnetospheric size. These smaller magnetospheres may result in radio emission originating at higher altitudes (and at larger corotation speeds, where relativistic aberration is enhanced), and covering larger solid angles than in the case of their younger counterparts~\cite{Ravi10}. In turn, this may lead to radio beams that are detectable at relatively larger impact angles $\beta = \zeta - \alpha,$ with $\alpha$ the inclination angle and $\zeta$ the observer angle, both being measured with respect to the rotation axis. MSPs therefore offer unique opportunities for studying pulsar emission geometries over a larger region in phase space than non-recycled, radio-loud $\gamma$-ray pulsars. In this context, it is interesting to note that all observed $\gamma$-ray MSPs are radio-loud. This is  in contrast to non-recycled pulsars. 
Given their much longer and more violent evolutionary history, including spin-up via accretion~\cite{Alpar82}, MSP magnetic fields may be more complex than those of younger pulsars. This may in part explain why their pulse profiles are more intricate and diverse. For example, MSPs were found to exhibit three classes of light curves (LCs) -- those where the radio leads the $\gamma$-ray profile, those where these are aligned, and those where the radio profile trails the $\gamma$-ray profile, termed `Class~I', `Class~II', and `Class~III' respectively. See \cite{V12}. In contrast, the radio leads the $\gamma$-ray profiles in the bulk of young pulsars (making them `Class~I'), and only the Crab shows phase alignment of radio and $\gamma$-ray profiles (`Class~II'). 
Lastly, MSPs generally seem to be more massive than their younger counterparts~\cite{Lattimer12} due to accretion of matter from their companion stars, which may enhance general relativistic effects such as frame dragging~\cite{MT92}. This may lead to increased electric fields which accelerate primary charges near the polar cap.

\section{Geometric modelling and light curve fitting}
We modelled the MSP radio and $\gamma$-ray LCs using standard emission geometries, including the outer gap (OG)~\cite{CHR86a}, two-pole caustic (TPC)~\cite{Dyks03}, altitude-limited OG and TPC (alOG, alTPC)~\cite{V12}, low-altitude slot gap (laSG)~\cite{V12}, and pair-starved polar cap (PSPC)~\cite{HUM05} geometric models. All these geometric models have physical counterparts\footnote{Technically, the PSPC model {\it is} the actual physical model in this case.} based on different electrodynamical assumptions of the acceleration. The latter lead to different locations and extents of acceleration regions (or `gaps') within the magnetosphere, associated with high-energy emitting regions. We combined these $\gamma$-ray geometries with a semi-empirical conal radio model~\cite{Kijak03,Story07}, but used a core component when indicated by polarimetry (e.g., when sense changing of circular polarisation is observed). 
Our models assume the retarded vacuum dipole magnetic field~\cite{Deutsch55} as the basic magnetospheric geometry, and correct for aberration and time-of-flight delays of emitted photons~\cite{Dyks04}, while assuming a constant emissivity in the corotating frame for all except the laSG and PSPC models. After collecting the emission (per solid angle) of a particular simulated pulsar in a 2D skymap ($\zeta$ vs.\ phase) for a fixed value of $\alpha$, we constructed LCs (by making constant-$\zeta$ cuts) on a grid of model parameters, and then used a maximum likelihood approach to obtain best joint radio / $\gamma$-ray LC fits for each 2PC MSP using {\it Fermi} LAT and radio LC data. Lastly, we estimated errors on the model parameters using either one- or two-dimensional likelihood profiles. More details may be found elsewhere~\cite{V12,V09,Johnson13}.

\section{Results: tentative trends}
\subsection{Towards a hybrid geometry?}
We find that no model can universally fit all MSP LCs. The OG and TPC models perform best, providing best fits for $83\%$ of the LCs ($\sim40\%$ each). The TPC geometry better fits LCs with significant off-peak $\gamma$-ray emission, while OG models prefer LCs with little or no off-peak emission. There are 28 Class~I MSPs, 16 being best fitted by the TPC, and 12 by OG geometry. There are 6 Class~II MSPs, 1 of which is best fitted by the alTPC model, 4 by the alOG model, and 1 by the laSG model. Lastly, the 6 Class~III MSPs are best fitted exclusively by the PSPC model. The above implies that a ``mix'' of models may be needed to fit all observed profiles, pointing to some hybrid model that incorporates features of the different models studied so far. Alternatively, a new geometry may be needed which may provide greater richness in terms of potential LCs it can produce, e.g., mimicking TPC LCs in some cases and OG LCs in other.

\begin{figure}[t]
\begin{center}
\includegraphics[width=18pc]{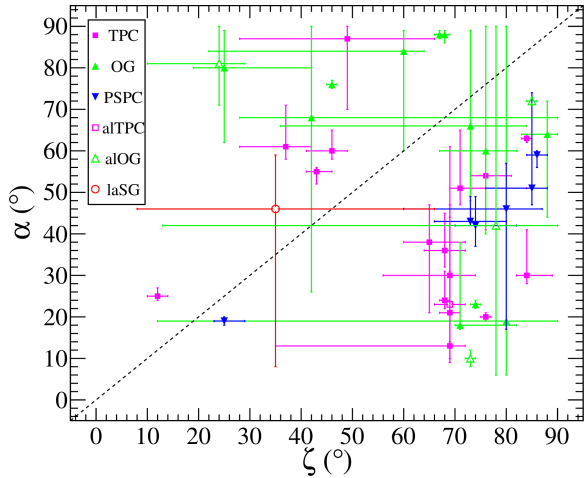}
\caption{\label{fig1} Best-fit $\alpha$ vs.\ $\zeta$. Different symbols distinguish different emission geometries.}
\end{center} 
\end{figure}

\subsection{Pulsar geometry and visibility constraints}
\label{sec:geom}
From our best fits of MSP LCs, we find a broad distribution of $\alpha$ (Figure~\ref{fig1}). This may argue against the idea~\cite{Young10} of spin-axis alignment with age, as the latter would imply a preference for smaller values of $\alpha$ for MSPs. This is in contrast with modelling results obtained for non-recycled $\gamma$-ray pulsars~\cite{Pierbattista13}, where  generally the best-fit $\alpha > 40^\circ$ (and young radio-loud pulsars have small $\beta$). The large range in $\alpha$ for MSPs partly reflects their wider radio beams and smaller magnetospheres which allow both the radio and $\gamma$-ray beams to be visible over a larger region of phase space than is the case for longer-period pulsars. While there may be younger pulsars with small $\alpha$ and large $\zeta$, their narrow radio beams may be missed due to an unfavourable observing geometry (i.e., large $\beta$). They will be radio-quiet $\gamma$-ray pulsars, if indeed $\gamma$-ray pulsations may be identified via, e.g., a blind periodic search~\cite{SazParkinson} (see Figure~7 of \cite{Pierbattista13}). The angle $\alpha$ for such pulsars may therefore be very uncertain, if pulsations are indeed found. Otherwise, these sources must be considered unidentified. The smaller range in $\alpha$ (and $\beta$) may thus be indicative of this selection effect. For the MSPs, on the other hand, a relatively larger range of $\beta$ is allowed, restricting the number of radio-quiet MSPs to very few. Indeed, there are no radio-quiet MSPs in the 2PC, while half of the young $\gamma$-ray pulsars are radio-quiet. 

There seems to be a `zone of avoidance' along $\alpha = \zeta$.  This was not expected {\it a priori} in terms of the possible $\alpha$ and $\zeta$ that MSPs may assume. While it would be interesting to see if this region is filled as more MSPs are discovered, this effect may point to some necessary refinement of our assumed radio emission geometry (i.e., one where profiles with lower multiplicities may be produced, even at $\alpha\approx\zeta$).
Figure~\ref{fig1} furthermore indicates a clustering at large $\zeta$, corresponding to the fact that it is generally the bright caustic emission (at large $\zeta$, near the spin equator) that is sampled by the observer to form the $\gamma$-ray profile peaks. A preference of $\zeta$ close to $90^\circ$ is also expected if pulsar spin axes are distributed randomly with respect to the Earth line of sight (in which case $\zeta$ should follow a $\sin\zeta$ distribution). 
The relatively larger $\alpha$ for the OG (vs.\ the TPC model) is connected to visibility -- the OG model is simply not visible when $\alpha$ is too small, due to the fact that it does not produce any emission below the null charge surface\footnote{Defined by the condition $\rho_{\rm GJ}=0$, where $\rho_{\rm GJ}$ is the Goldreich-Julian charge density~\cite{GJ69}.} as in the case of the TPC model. In the PSPC case, the majority of best-fit $\alpha$ values lie in the range $40^\circ - 60^\circ$, giving optimal off-peak emission levels and radio peak multiplicities. Interestingly, all currently modelled Class~III MSPs (using the PSPC model) have $\zeta > \alpha$.

\begin{figure}[t]
\begin{center}
\includegraphics[width=18pc]{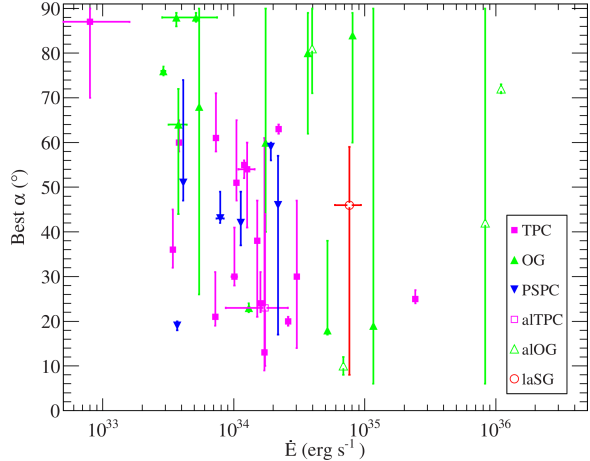}
\caption{\label{fig2}Plot of best-fit $\alpha$ vs.\ $\dot{E}$.}
\end{center}
\end{figure}

\subsection{An evolutionary trend?}
Figure~\ref{fig2} indicates the best-fit $\alpha$ vs.\ spin-down power $\dot{E}$. The TPC model best-fit results hint at an inverse trend between $\alpha$ and $\log_{10}\dot{E}$ for Class~I MSPs (the Pearson correlation coefficient is -0.47 and the chance probability is $1.3\times10^{-2}$). At larger $\dot{E}\propto P^{-3}$, we have smaller $P$, so that smaller values of $\alpha$ seem to be associated with shorter-period pulsars in this case. If we accept that $\zeta$ is generally large (Section~\ref{sec:geom}), these results may imply a larger impact angle $\beta$ for these pulsars. This corresponds to the fact that the TPC is indeed visible for larger values of $\beta$ when compared with the OG model, since it includes low-altitude emission that is not present in the OG geometry. Furthermore, since all MSPs in this plot are radio-loud, we expect that the radio beams of high-$\dot{E}$ MSPs will generally be at higher altitudes. These would therefore be wider, allowing one to probe smaller $\alpha$. Lower-$\dot{E}$ MSPs may not be visible in radio if their narrower radio beams are pointing away from the observer and so may not be easily identified as $\gamma$-ray pulsars. On the other hand, there may be hints of an underlying evolution of $\alpha$ toward the equator with age (smaller values of $\dot{E}$). A population synthesis approach would be necessary to disentangle the effects of visibility and obliquity evolution. In any case, noting the presence of outliers and points with large uncertainties, this observed trend should be regarded with caution.

\subsection{Caustic radio emission}
In the case of Class~II MSPs, we find that the radio emission may be caustic in nature (i.e., emission originating at different altitudes in the magnetosphere being bunched in phase by relativistic effects; see~\cite{Morini83}), since radio and $\gamma$-ray profiles are phase-aligned, implying a common origin of the emission. Radio emission would therefore originate at higher altitudes. This is in contrast with the usual low-altitude conal emission found in the Class~I and Class~III MSPs~\cite{V12}, and should be expected to be associated with depolarisation and rapid position angle swings, since the emission from a large range of altitudes and magnetic field orientations is compressed into a narrow phase interval when forming the bright peaks~\cite{Dyks04}. Although the radio emission altitudes cannot be well constrained by current statistics, we do find that the radio and $\gamma$-ray emission regions typically have significant overlap, while the radio is generally more limited in altitude and originates higher up than the $\gamma$-rays.

\subsection{Tapping the power source -- luminosity and beaming}
\begin{figure}[t]
\begin{center}
\includegraphics[width=20pc]{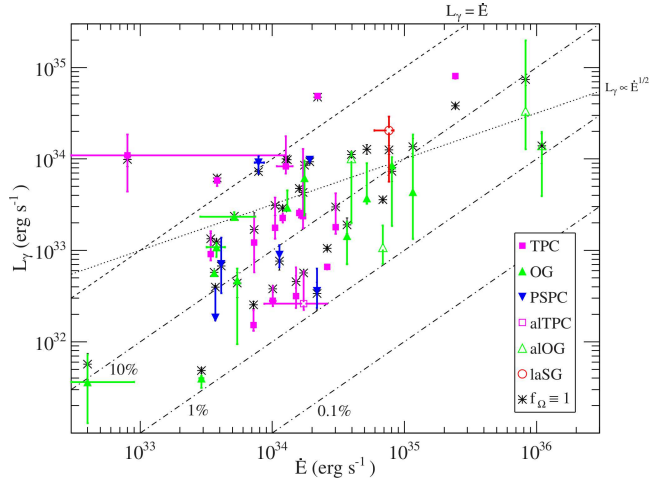}
\caption{\label{fig4}Plot of $L_\gamma$ vs.\ $\dot{E}$. Black stars indicate values where $f_\Omega=1$ was assumed. The dashed and dash-dotted lines show efficiencies of 0.1\%, 1\%, 10\%, and 100\%. The dotted line indicates $L_\gamma \propto \dot{E}^{0.5}$ with arbitrary normalisation.}
\end{center}
\end{figure}

The $\gamma$-ray luminosity $L_\gamma$ is a very important intrinsic parameter characterising how rotational energy is converted into $\gamma$-ray emission. This is estimated from the observed $\gamma$-ray energy flux $G$ using $L_\gamma = 4\pi f_\Omega G d^2$ (e.g., \cite{Watters09}), where $f_\Omega$ is the beaming correction factor which accounts for the fact that emission is not beamed isotropically. We can estimate $f_\Omega$ from the emission pattern implied by the model by comparing the total emitted pattern to the one observed at a particular $\zeta$ for given $\alpha$. Our fits indicate that $f_\Omega < 1$ in most cases. We are thus typically sampling emission that is above the average emission level. Conversely, for PSPC best-fit profiles, we have $f_\Omega > 1$ in most cases, since we are missing the brightest part of the emission concentrated at low altitudes near the polar caps. The evolution of $L_\gamma$ with $\dot{E}$ is very important, as this characterises the regime in which energy conversion takes place. It is expected that younger pulsars find themselves in a screened-potential regime, characterised by a relation $L_\gamma\propto I_{\rm PC}\propto\dot{E}^{1/2}$ (where $I_{\rm PC}$ is the polar cap current, and the polar cap potential $V_{\rm PC}$ is roughly constant in this case), while older pulsars operate in a rather more pair-starved regime where conversion of emitted $\gamma$-rays into electron-positron pairs is inefficient, and $L_\gamma\propto V_{\rm PC}I_{\rm PC}\propto\dot{E}$~\cite{HMZ02}. We find that $L_\gamma$ roughly follows a linear trend with $\dot{E}$ for the MSPs, which is consistent with this expectation (Figure~\ref{fig4}). 
Lastly, we observe a clustering around a $\gamma$-ray efficiency of $L_\gamma/\dot{E} = 10\%$.

\subsection{Discriminating between the different LC classes}
\begin{figure}[t]
\begin{center}
\includegraphics[width=17pc]{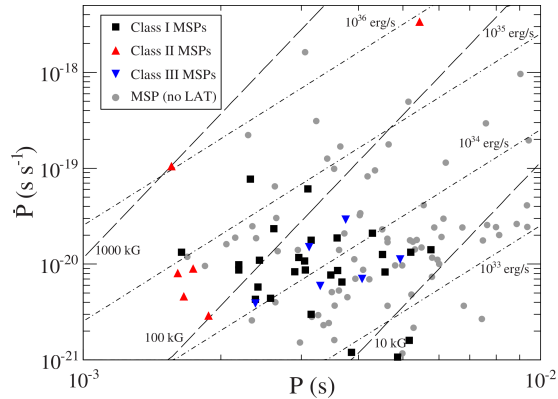}
\caption{\label{fig3}The $P\dot{P}$ diagram indicating different MSP LC classes. Contours of constant $\dot{E}$ and $B_{\rm LC}$ are also shown.}
\end{center}
\end{figure}

Figure~\ref{fig3} shows the positions of the modelled MSPs on a period-period-derivative ($P\dot{P}$) diagram, with the different classes differentiated by different symbols as described in the legend. Grey dots are radio MSPs with no detection in 2PC. Contours of constant magnetic field $B_{\rm LC}$ at the light cylinder are indicated by dashed lines, while constant-$\dot{E}$ contours are indicated by dot-dashed lines, assuming dipole spin-down and a canonical moment of inertia of $I=10^{45}$~g~cm$^2$. We see no clear differentiation of MSP LC classes according to the usual pulsar variables such as $P$, $\dot{P}$, $\dot{E}$, or $B_{\rm LC}$, although it seems that Class~II MSPs favour lower values of $P$, and Class~III MSPs scatter about low values of $B_{\rm LC}$.

\section{Conclusion}
We described our attempt to model the LCs of all MSPs that appear in {\it Fermi} LAT's 2PC. We noted some tentative trends (e.g., a broad distribution in $\alpha$, clustering at large $\zeta$, and a linear relation between $\log_{10}L_\gamma$ vs.\ $\log_{10}\dot{E}$), which may strengthen as more data are accumulated. A new hybrid model may be needed to unify the different older models and reasonably reproduce all existing LCs. Different LCs classes are not easily distinguished based on canonical pulsar variables alone, but may rather be a reflection of the complex electrodynamical environment of the pulsar. We intend to next study the effects of new magnetic field geometries (e.g.~\cite{HM11, Breed13, Kalapotharakos13}) and more complex higher-altitude radio emission patterns on the predicted MSP LCs.
\ack
\small{CV is supported by the South African National Research Foundation. AKH acknowledges support from the NASA Astrophysics Theory Program.  CV, TJJ, and AKH acknowledge support from the \textit{Fermi} Guest Investigator Program.
The $Fermi$ LAT Collaboration acknowledges support from a number of agencies and institutes for both development and the operation of the LAT as well as scientific data analysis. These include NASA and DOE in the United States, CEA/Irfu and IN2P3/CNRS in France, ASI and INFN in Italy, MEXT, KEK, and JAXA in Japan, and the K.~A.~Wallenberg Foundation, the Swedish Research Council and the National Space Board in Sweden. Additional support from INAF in Italy and CNES in France for science analysis during the operations phase is also gratefully acknowledged.}

\end{document}